# A Systematic Security Evaluation of Android's Multi-User Framework

Paul Ratazzi*†, Yousra Aafer†, Amit Ahlawat†, Hao Hao†, Yifei Wang† and Wenliang Du†

*Information Directorate, Air Force Research Laboratory, Rome, NY
†Dept. of Electrical Engineering & Computer Science, Syracuse University, Syracuse, NY

*Abstract*—Like many desktop operating systems in the 1990s, Android is now in the process of including support for multi-user scenarios. Because these scenarios introduce new threats to the system, we should have an understanding of how well the system design addresses them. Since the security implications of multi-user support are truly pervasive, we developed a systematic approach to studying the system and identifying problems. Unlike other approaches that focus on specific attacks or threat models, ours systematically identifies critical places where access controls are not present or do not properly identify the subject and object of a decision. Finding these places gives us insight into hypothetical attacks that could result, and allows us to design specific experiments to test our hypothesis.

Following an overview of the new features and their implementation, we describe our methodology, present a partial list of our most interesting hypotheses, and describe the experiments we used to test them. Our findings indicate that the current system only partially addresses the new threats, leaving the door open to a number of significant vulnerabilities and privacy issues. Our findings span a spectrum of root causes, from simple oversights, all the way to major system design problems. We conclude that there is still a long way to go before the system can be used in anything more than the most casual of sharing environments.

## I. INTRODUCTION

As the Android operating system evolves, and the devices supporting it become more capable, more advanced functionality and features become available to the end-user. Two of the most recent major enhancements to Android are multiple users (MU), introduced in version 4.2 (API 17) in November 2012, and restricted profiles (RP), introduced in version 4.3 (API 18) in July 2013. Targeted towards sharable devices such as tablets, these enhancements strive to provide individual user spaces on a single physical device. Each user space supports a separate set of accounts, apps, settings, files, and user data, distinct from those of the primary owner [1]. Google has introduced two means of adapting Android to address these multi-user scenarios.

*Multiple Users* (MU) designates the main account as *Owner*. Through the device settings, the owner account may create additional MU accounts. These secondary accounts are essentially the same as the owner, except for the fact that they cannot manage (i.e., create, modify, delete) other users. MU accounts enjoy most of the other privileges and functionality of the owner, including managing the device's wireless and network settings, pairing Bluetooth devices, customizing sound and display settings, installing/removing their own apps, adjusting privacy settings (e.g., location access), and configuring security features (e.g., screen lock, credentials). Each account also has a separate virtual SD card storage area within the physical SD card.

*Restricted Profiles* (RP) are similar to MU accounts, but they lack several key functionalities compared to owner and MU accounts. Like MU accounts, RP accounts cannot manage users. In addition, RP accounts are restricted from installing apps. Instead, the owner account "turns on" specific apps from the set of installed apps for the RP account.

Although one might reasonably assume that these enhancements would provide isolation among users and profiles similar to that provided by today's desktop multi-user systems, forebodingly, Google's own recommendation regarding their use is for owners to share their device with only people they trust [2]. To security-minded individuals, statements like these raise red flags. In our case, it gave extra motivation to our investigation.

### A. Motivation

Since its introduction, the Android operating system has enjoyed tremendous success, to the point now where over 1.5 million new devices are being activated *every day* [3]. This astounding growth is not only in regards to the sheer number of devices running Android, but also in terms of the different types of devices running the system. More so than ever, Android is expanding beyond the smartphone and becoming the operating system of choice for a variety of "keyboardless" and embedded devices such as tablets, home entertainment equipment, automobile dashboards, and appliances. While smartphones are usually personal single-user devices, many of these other applications and devices exist in a multi-user environment. Thus, the need for multi-user support in Android grew along with its expansion from strictly personal devices to those with varied purposes in multi-user environments. Today, upwards of 27% of devices in use are running one of the three newest versions of the operating system capable of the multi-user features that debuted a mere 18 months ago. It is clear that Android is rapidly *evolving* towards a multi-user environment, rather than having it designed-in from the beginning, a path which is concerning from a security point of view given the clear and substantial implications that a multi-user environment has on the system.

This uncertain foundation and rapid expansion of Android into multi-user environments provides the motivation and impact of our work. By gaining insights through detailed understanding and systematic exploration of the system, we are able to hypothesize about potential security problems and design effective experiments to test them. Our findings indicate that the evolution towards a system that meets established security principles is far from complete, and point to a failure to reconsider a change in the original assumption of a single benign user. We demonstrate situations whereby a secondary user can bypass restrictions, gain unauthorized privilege, spy on other users, and create denial of service. Our findings represent a spectrum of root causes, from what we believe are simple oversights, all the way to major system design problems.

*B. Threat Model*

As stated earlier, Google recommends sharing multi-user devices only with trustworthy people. Unfortunately, varying definitions of trust, different expectations for security and privacy, and a wide variety of use cases make this a very ambiguous statement. On the other hand, immediately identifying a specific threat model or scenario at the outset of our investigation would risk narrowing the field of potential insights we are hoping to gain. Instead, we have deliberately designed our investigation such that insights and knowledge are first gained through systematic analysis, independent of any particular threat mindset. Only then do we pose hypotheses that factor in specific threat scenarios. Details of these are given in Section III-C.

The rest of the paper is organized as follows: Section II gives more details on how multi-user has been implemented in Android. Section III describes how we went about discovering and evaluating key aspects of the new system features. Section IV presents our findings, Section V gives an overview of related work, and Section VI discusses the findings and concludes the paper.

## II. BACKGROUND

Before describing our investigation, a brief technical overview of the implementation of multiple users is needed. The following section is divided into four parts: Android framework extensions, filesystem configuration, kernel mechanisms, and run-time considerations. For this discussion, the Linux user ID and group ID are referred to as `uid` and `gid`, respectively, while IDs within the Android framework are denoted by `userId`, and `appId`.

*A. Framework -* `userId`

Version 4.2 added the `android.os.UserHandle` class to represent multiple users on the device. This class designates `userId` 0 as the device owner, and several special `userIds` to represent all users (-1), the current user (-2), the current user or self (-3) and the null user (-10000). Actual `userIds` are assigned by the `UserManagerService` (also introduced in 4.2) when new users or new restricted profiles are created by the owner. This class defines the starting `userId` as 10, and increments it by 1 every time a new user is created, until the number of current users equals the maximum number defined by the system property `fw.max_users`. State is maintained in `/data/system/users/userlist.xml`, where a list of currently-assigned users and the next available `userId` is stored. `userIds` are not re-used when a user is deleted, in order to avoid leaking an old user's files to a new user. `userIds` are assigned in the same way regardless if they are for a secondary user or a restricted profile.

As has always been the case, each installed application is assigned an `appId`.[1] `android.os.Process` class defines ranges of `appIds` that can be assigned to different types of apps. Normally, these IDs range from 10000 to 99999.

In the past, there was no defined `userId`, so the `uid` was just the same as the `appId`. To enable kernel isolation of apps among multiple users and profiles, the `uid` is obtained by combining `userId` with `appId` using the expression `uid = userId * PER_USER_RANGE + (appId % PER_USER_RANGE)`, where the default `PER_USER_RANGE` is 100000. Likewise, `userId` and `appId` can be found from `uid` with `userId = uid / PER_USER_RANGE` and `appId = uid % PER_USER_RANGE`, respectively. The new `UserHandle` class includes the methods for performing these conversions.

Thus, `uid` is a two-digit `userId` (00, 10, 11, 12, ...) concatenated with a five-digit `appId` (10000, 10001, ...). For example, an app with `appId` 10056 will run with `uid` 0010056 when started by the owner (`userId` 0), and `uid` 1010056 when started by the first secondary user or restricted profile (`userId` 10).

System `uids` not directly associated with apps are still in the range 0-9999. For example, `root` is `uid` 0, `system` is `uid` 1000, `radio` is `uid` 1001, and `shell` is `uid` 2000.

*B. Framework - Permissions*

Several new permissions have been introduced with the advent of multi-user support. These include `MANAGE_USERS`, `INTERACT_ACROSS_USERS` and `INTERACT_ACROSS_USERS_FULL`, which are used to protect some types of inter-user functionalities such as `startActivityAsUser()`. Generally, checks for these permissions are bypassed if the calling process has a `root` or `system uid`. Several are also bypassed for processes running as `shell`. As `signatureOrSystem` permissions, they will not be granted to apps not in the `/system` partition or signed with the platform key.

*C. Framework - Package Management*

To accommodate multiple users, Android's package management system was modified so that secondary users can choose different sets of installed apps, and the owner can choose which apps are enabled or disabled for RPs. However, as currently implemented, package management is still largely *platform*-centric rather than *user*-centric. Although it may

---
[1]Before the introduction of multi-user, `uid` and `userId` were used interchangeably to refer to the unique identifier for each app installed on the system. In versions with multi-user extensions, `userId` is used to denote the actual user, while `appId` is the designation for each app's unique ID. However, there are still several instances of code and files that use `userId` to refer to apps. For example, the `sharedUserId` tag in `AndroidManifest.xml` actually refers to package names which will share the same `appId`.

*appear* that each user has their own independent set of apps installed, in reality, each app is installed *once* for the entire platform, and then either enabled or disabled for each user. Evidence of this is seen in the fact that a device with multiple users still has only one `packages.list` file to map package names to corresponding data directories, `appIds`, and `gids`. Moreover, there is only one `packages.xml` file for associating package names and `appIds` with signature keys, native library paths, code paths, granted permission(s), and special conditions such as `sharedUserIds`. Notably, this file also associates the aggregate permission list for `sharedUserId` packages to `appId` without regards to any particular `userId`.

Because the content and structure of the above files was not changed for multiple users, `PackageManager` keeps track of each user's specific app installation status by way of a `package-restrictions.xml` file for each user (default location: `/data/system/users/<userId>/`). As such, when an individual user installs an app for themselves, the app is really installed for all users on the entire platform, and then simply "hidden" from other users by tagging the package name with `inst=false` in `package-restrictions.xml`. We confirmed this by manually removing a `<pkg name>` tag that contained `inst=false` from a secondary user's `package-restrictions.xml`, and observing that the app is now available to that user.

### D. Filesystem

To support multi-user, several changes to the filesystem organization were made. Whereas the single user's app data was previously stored under `/data/data/<package_name>`, this data is now isolated for each user under `/data/user/<userId>/<package>`. To maintain backwards compatibility, the owner's (`userId` 0) app data is still stored under `/data/data/<package>`, with a symbolic link from `/data/user/0` to `/data/data`. Subdirectories in these locations are owned by the `uid` for the respective user and app. Strong isolation is achieved through the use of Linux bind mounts and filesystem namespaces [4].

### E. Kernel

Since Linux is naturally a multi-user system, implementing Android's multi-user extensions at the kernel level did not require any changes to the kernel itself. For all versions of the Linux kernel used in Android, the Linux `uid` is an unsigned 32-bit integer which can represent over 4 billion unique `uids`. Thus, the `uid` discussed above, formed from the Android `userId` and `appId`, uniquely identifies both the user and app, and is directly used as the Linux `uid`. In this way, standard Linux discretionary access control (DAC) can provide isolation not only among apps, but also among each user's installation of a particular app.

### F. Run-time

On a running device, only one user can be "logged in" at any one time. However, through the switch users function, multiple users introduced the concept of the *current user*, which refers to the user interacting with the device. We refer to other users who may have been using the device before it was switched to the current user as "inactive users." Although these users cannot interact with the device, many of the underlying processes associated with their active session, are left running. Their apps are paused and their background services may be left to run. On builds we have used, there is a limit of 3 to the number of users that can be inactive before their processes are completely removed.

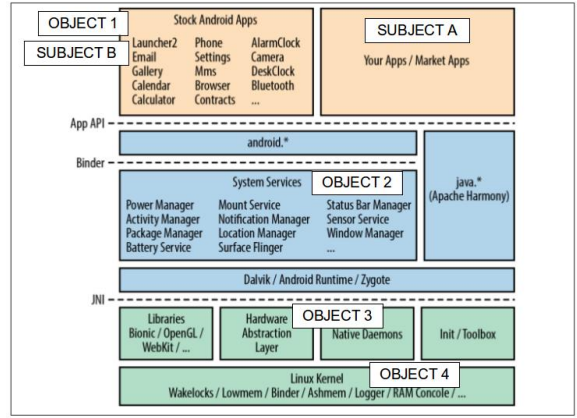

Fig. 1. Investigation problem space showing various subject-object combinations. Adapted from [5] with the permission of O'Reilly Media, Inc.

### III. INVESTIGATION METHODOLOGY

At the core of any security investigation lies the question of whether the system design is based on valid assumptions. As Android evolves into a multi-user system, what once may have been a set of valid assumptions may suddenly be undermined by emerging system characteristics and/or usage models. In particular, the original assumption of a benign, single-user environment is no longer valid. Rather than a single owner who has administrative authority over most aspects of system configuration and would not attack or intentionally mis-configure his own system, there is now an environment where it is plausible for secondary users to bypass restrictions, attack other users, or deliberately reconfigure the system in an unauthorized way.

### A. Scope

In defining the scope of our investigation, we look towards an overall architecture of Android. We begin with Yaghmour's [5] high-level architectural view shown in Fig. 1. Although this diagram is not specific to multi-user, we consider it in that context because of our focus. This diagram shows broad categories of system resources such as stock apps and system services, which become our subject and object categories. Our goal is to find subject-object combinations that have interesting security aspects unique to the multi-user case, and then evaluate the suitability of the access control path(s) between them. We specifically focus on scenarios whereby a secondary user exercises all possible paths to access resources and/or gain privileges.

*1) Subjects:* With the above in mind, the subjects we consider are apps and user interface (UI) elements that the secondary user can install and use. This list includes user-installed as well as stock apps (e.g., Settings), with the key difference being privilege (stock apps can have `signatureOrSystem` permissions while user-installed apps cannot). In Fig. 1, these are labeled as SUBJECT A for the case of a user-installed app making API requests to other parts of the system, and SUBJECT B for cases whereby the user interacts with stock apps to access resources.

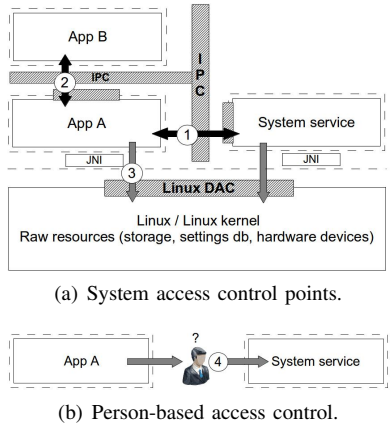

(a) System access control points.

(b) Person-based access control.

Fig. 2. Simplified access control models.

A very important difference about the subjects we consider compared to those of the single-user case is that the secondary user who launched the app may not necessarily still be the current user. For example, components of an app launched by a secondary user prior to switching to another user may still function.

*2) Objects:* The resources available to the subjects discussed above are represented as the object in our access control investigation. Examples from Fig. 1 include public interfaces of other apps (OBJECT 1), services (OBJECT 2), abstracted hardware devices (OBJECT 3) and kernel objects (OBJECT 4).

The most important difference about objects we consider compared to those of the single-user case is that some resources may be shared with other users on the device. For example, hardware devices such as the camera, or common settings databases are objects that are shared among multiple users on the device.

*3) Access control paths:* Between these subjects and objects are communication paths that may include access control mechanisms. We draw upon the work of [6], [7], and [8] to present the simplified models of Fig. 2 that show communication paths pertinent to our investigation.

Fig. 2(a) depicts each app and system service contained by separate sandboxes as indicated by the dotted lines around them. Communication among these sandboxes (denoted by bi-directional arrows) is done through Intents and Binders. Using Intents, apps may request access to services or providers (path ①) or launch exported activities of other apps (path ②). These paths include access control points provided either by the system (as part of the Intent mechanism), or at the public interface of the object itself. Using the native interface, apps may also use system calls to directly request resources controlled by the Linux kernel (path ③), and these are subject to Linux DAC. These three paths are permission-based, access control list (ACL)-based, or some combination of these.

Fig. 2(b) shows another communication path with access control typical of smartphones and tablets, that performed by the user (path ④). In this case, the current user makes the decision to allow or disallow access to a resource such as location, for instance. We refer to this as *person-based* access control to avoid confusion with the notion of users on the device.

A fifth type of path, not shown, are those that have *no* access control along them.

*B. Questions & Insights*

As we study the inner-workings of Android's multi-user features, we are able to make two observations. First, the new features have introduced important new considerations for the subjects and objects shown in Fig. 1. Examples of this include the concept of apps run by a userId that is different than that of the current user, and person-based access control decisions being made by multiple individuals. Second, even though none of the access control paths of Fig. 2 are unique or dedicated to the extensions, some have been modified to account for the presence of multiple users on the device. Examples include methods that include checks for INTERACT_ACROSS_USERS permission and apps that express different versions of their UI to restricted users than they do to the device owner.

These observations lead us to the following top-level questions for our investigation:

1) Do Android's access control points properly account for the new considerations regarding subjects and objects?
2) If not, can a secondary user exploit these shortcomings, and what is the potential damage?

In order to answer these questions, we enumerate all of the meaningful subject-object combinations within the broad categories identified by Fig. 1, and identify the corresponding access control paths from Fig. 2. This gives us a comprehensive list of specific things to study. For example, a user-installed app (SUBJECT A) can send an Intent using startActivity() to launch any exported activity of any other app (OBJECT 1). Thus, we study the system's Intent mechanism and the specifics of how these activities are exported. Specifically, we examine the source code in order to determine what considerations, if any, do the Intent mechanisms and exported activities give to multiple users. If none or partial, we consider whether there should be protections and how a secondary user might exploit the shortcomings.

*C. Hypotheses About Multi-User Security*

This last step allows us to develop a set of hypotheses which can be used to design experiments for testing the adequacy of access controls and demonstrating the consequences. We present a partial list of our most interesting hypotheses here:

*1) Secondary users may be able to bypass their restrictions by exploiting the unprotected public interfaces of system apps:* Secondary users are supposed to lack certain capabilities that the owner has, such as mobile plan settings. However, from our study of how access control restrictions are implemented

in Settings, we see that many are accomplished by way of hiding portions of the UI, while the corresponding activities are exported publicly. This situation corresponds to a particular OBJECT 1 in Fig. 1 (Settings) that is shared among all users without adequate access control along path ② of Fig. 2(a). Results from testing this hypothesis are contained in Section IV-A.

*2) Secondary users may be able to maliciously reconfigure critical platform-wide settings that are persistent across user switches:* Secondary users possess certain administrative capabilities (e.g., network settings) that are normally reserved for privileged users on mature multi-user systems such as Linux. Under Android's single user assumption, some of these settings are protected by person-based access control since UI interaction by the benign user is required to prevent malicious apps from making invisible changes programmatically. However, when the benign user assumption is invalid, Fig. 1's shared resources protected only by Fig. 2(b)'s person-based access controls (path ④) can be maliciously manipulated. The consequences are even more severe for cases where the configurations are persistent across user switches, such as in the case of network configuration. Results from testing this hypothesis are contained in Section IV-B.

*3) Inactive users may be able to spy on active users by exploiting improper access control enforcement on shared hardware resources:* As mentioned above, multi-user extensions introduce the concept of current and background users. However, unlike a true multi-user system such as Linux, which generally allows multiple remote logins simultaneously, there can only be one active user "logged in" an Android tablet at any one time. However, our enumeration of Fig. 1 objects discovered apps and services that have access to shared objects and are allowed to continue operations even after a user switch occurs. Of these, certain ones such as audio, camera and location have obvious privacy implications if used without the current user's knowledge or consent. From an analysis of the implicated access control paths ① and ④ in Fig. 2, we find that authorizations granted when the secondary user is the current user may not be properly reconsidered after a user switch. Results from testing this hypothesis are contained in Section IV-C.

*4) `sharedUserId` permissions may not be properly separated when `sharedUserId` apps are installed by different users:* Multiple users extensions bring with them the idea that each user may have different settings, preferences, and apps. Obviously, these should be isolated such that one user cannot accidentally inherit permissions or capabilities from another. Because our enumeration of all subject and object combinations of Fig. 1 included apps that leverage the `sharedUserId` feature, we discovered problematic situations with overprivilege that can occur when different users install apps with `sharedUserId`s. In particular, we see that access controls at Fig. 2(a) locations ①, ②, and ③ fail to differentiate the subject because each `sharedUserId` app's permissions are commingled with others of the same `appId` in a single `packages.xml` file shared among all users. Results from testing this hypothesis are contained in Section IV-D.

*5) A malicious user may be able to exploit the shared package management system to modify another user's app bytecode or prevent them from installing apps with package names identical to ones installed by the attacker:* The shared package management mechanism that led to Hypothesis 4 is also the cause of other problems. Since package installation is platform-centric rather than user-centric, changes by any user authorized to install apps will affect all users on the platform. Specifically, if a secondary user upgrades a package, the bytecode changes affect all users that have that package installed. Likewise, if a malicious user installs a fake app with a real app's package name, all users are prevented from installing the real app. Results from testing this hypothesis are contained in Section IV-D.

To test these hypotheses, we designed and conducted experiments using Android `4.4.2_r1` [9]. The details of these experiments and our findings are the subject of Section IV.

## IV. Findings

### A. Unprotected Activities

Hypothesis 1 states that secondary users may be able to bypass their restrictions by exploiting the unprotected public interfaces of system apps. To find out if this is true, our experiment must first identify the intended restrictions placed on a secondary user, and then compare them with the full set of exposed interfaces.

To understand the *intended* restrictions on secondary users, we systematically mapped and compared the UI accessible to the owner with that for a secondary user. A privileged app where significant differences have been observed is Settings. Settings is important to consider from a security point of view because it is granted the `SignatureOrSystem` permissions such as `WRITE_SECURE_SETTINGS`.

From our UI mapping, we observed that Settings implements a number of UI restrictions based on type of user by hiding certain menu items. As such, we assume these are capabilities that secondary users are not supposed to have. As an example, we consider virtual private network (VPN) settings which is hidden from the secondary user by way of logic within `WirelessSettings.java`. This logic compares the current user's `userId` with that of the owner and executes `removePreference(KEY_VPN_SETTINGS)` if not equal.

With an understanding of how Settings presents a restricted UI to secondary users, we compared the list of restricted UI elements with exported activities in the app's manifest to find which of these elements can be launched directly via intent [10]. Among this set is our VPN example, which can be accessed by secondary users with the following code:

```
Intent intent = new Intent();
intent.setClassName("com.android.settings",
    "com.android.settings.Settings$VpnSettingsActivity")
    ;
startActivity(intent);
```

Among the many other examples we found are mobile network & mobile plan settings (under Wireless & Network settings), and backup & reset settings (under Personal settings). Secondary users can access these activities because of a lack of access control along path ② of Fig. 2(a), such as a check of

`UserHandle.myUserId()`. Thus, each of these examples represent potentially dangerous situations since these activities allow a secondary user to manipulate configuration settings that may be able to be used to negatively affect the owner or other users of the device.

As it turns out, our example of VPN contains additional access control checks in `Vpn.java` that do properly identify the subject and prevent restricted users from connecting VPNs. Thus, for VPNs at least, the hypothesis is only partially true. Because of the numerous cases of restricted UI elements also being exported to all users, we intend to further investigate these other cases in our future work.

### B. Unrestricted Administrative Functions

Hypothesis 2 states that secondary users may be able to use device configurations which are persistent across user switches to attack other users. Although related to Hypothesis 1, this case does not involve a user bypassing restrictions, but simply implementing a malicious environment using the UI elements freely available to them.

To test this hypothesis, we built an experiment around network configuration, since this function is usually reserved for administrative users on standard multi-user platforms. We found that all users, secondary, restricted profile or otherwise, have full access to WiFi settings and can add and configure network connections as they choose. Furthermore, these settings are common to all users since they are ultimately stored by the system in a single `/data/misc/wifi/wpa_supplicant.conf` file that has no provisions for identifying the user who has authorized a particular connection. Finally, our experiment showed that WiFi connections are persistent across user switching.

This arrangement enables a secondary user to connect a multi-user device to a malicious hotspot and control all traffic to/from the device while it is being operated by other users. The hypothesis is true and the situation represents the fifth case mentioned in the Fig. 2 discussion, that of *no* access control.

### C. Use of Sensors and Hardware Devices by Multiple Users

Android provides various hardware features such as media resources and sensors. In single user context, all hardware interfaces belong to the same user without any constraints. Ideally, in multi-user context, a hardware resource should only be bound to a single user at a time, defined by the currently logged in user. Since the hardware interfaces are shared among the users on an Android device, the transition from single to multiple user framework requires changing the access control model on hardware resources to make sure that a hardware resource is only granted to the logged in user.

In this section, we aim to answer Hypothesis 3 from Section III-C. That is, non-logged in secondary users can exploit improper access control enforcement on shared hardware resources to spy on current users. In fact, if Android does not enforce proper access control on shared hardware resources based on user status, a non-current user can still use a hardware interface to infer various information about the logged in users and spy on them. If a non-current user can query the light and accelerometer sensors over a time interval, he can infer potential activities about current user such as whether he is sitting indoors, or jogging outdoors. Moreover, if he can query the GPS service, he can even infer where he is sitting or jogging. Even more concerning, if he can launch the sound and camera recorders, he can know easily more details such as with whom he is and what type of conversation he is having.

To ensure that a hardware resource is only bound to the currently logged in user, Android should be able to identify if the user requesting a resource is logged in. Also, it should track if the user who initiated the request is continuously logged in during the service lifetime. More specifically, if a user switching happens, Android should be able to revoke any resource access from non-logged in users.

Our investigation focuses on checking if Android multi-user framework enforces access control on shared hardware resources based on user status. We systematically study media resources and common sensors; for each studied resource, we design an attacking app that will enable a user to try to access that resource even if he is not logged in. We exploit the fact that `ActivityManager` does not kill all non-current user processes unless under limited memory usage to create an attacking channel and potentially leak information from hardware resources about a victim user. The attacking app is launched when the malicious user is logged in, and will be continuously running after he logs out and the victim user logs in. We specifically assigned a non-owner user to be the attacker, and the owner to be the victim since non-owner is less privileged compared to the owner. We report our findings below.

*1) Media resources:* Android media resources provide various media interfaces such as video and audio recording and playing. Ideally, to ensure that a non-logged in user is not able to use its services, the following two access control points should be enforced:

- At request time: when an app requests a resource, the system should check if its `userId` is equal to the current `userId`. If not, the request should fail.
- At user switch time: when a user is logging out, the system should revoke any earlier resource access and stop the recording.

To check if these access control points are properly enforced, we designed an app that launches the camera (without a preview window) and starts video recording under two scenarios:

- The app intentionally schedules video recording when the victim user is logged in (in an estimated amount of time).
- The app starts video recording immediately while the attacker is logged in.

We launch this app from the attacker user account (`userId` 10) and switch to the victim account (`userId` 0). In the two scenarios, our app is able to record video in the context of the victim account. The recordings will be saved under the non-current attacker's app data directory.

Fig. 3(a) demonstrates the process of our attack. The magnified icon on the attacker screen indicates the running camera recording app. After user switch, the recording app is not killed and continues working in the background. Fig.

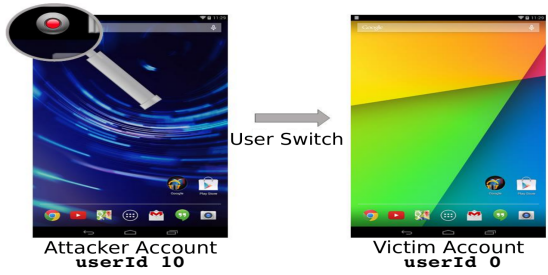

(a) Spying attack mechanism.

(b) Process list showing attacker's process (`u10_a76`) still running while victim (`userId 0`) is active.

Fig. 3. Exploiting shared media resources to spy on current users.

3(b) shows the output of the Linux `ps` command at the same time as the victim screen in Fig. 3(a) is being shown. The red rectangle highlights the attacker's process, left running even though the attacker is no longer the current user. Moreover, as depicted on the victim screen, there is no icon or notification showing the background recording app. The victim is thus not aware that he or she is being recorded, unless she checks the currently running apps belonging to the other user through the Settings app or by running `ps` as we have done. In either case, the victim does not have the privilege necessary to stop the process, short of powering off the device.

The success of our attacking app under the first scenario reveals that there is not proper access control at request time based on user status. To confirm this conclusion, we investigate how an app can launch media recording. In fact, the app has to invoke the method `start()` on the `MediaRecorder` class. This class initiates an interprocess communication (IPC) call to the `MediaServer`, which provides all services related to Androids Media framework such as media playback and media recording. Ideally, on `MediaServer`'s side, Android has to check if the user requesting to start the recording is currently logged in and grant the service accordingly. However, `MediaServer` does not perform this access control, leading to the success of our app under the first scenario.

By inspecting the call chain of `MediaRecorderClient::start()` within `MediaServer`, we observer that there is no access control based on user status at any point. Even if the access control is correctly granted at request time based on the status of the user, `MediaServer` should also be able to revoke the granting once a user switch takes place, since it will invalidate the earlier access control decision. However, since our app succeeds under the second scenario (i.e., does not stop recording when user switch happens), we can infer that `MediaServer` does not perform any access control at switch time based on user status. In fact, based on our inspection of the `MediaServer` code, we found that user switch is not handled at all.

*2) Motion, environmental and position sensors:* Most Android devices have built-in sensors that provide various information such as motion, environmental, and position. Motion sensors include accelerometers, gravity, rotational vector sensors, and gyroscopes. Environmental sensors measure ambient air temperature, pressure, illumination and humidity, while position sensors measure the physical position of a device.

Unlike the media recording activity, these sensors' activity follows an event-driven approach. That is, an app first registers a listener to receive sensor events through the `SensorManager`, then the `SensorService` will deliver any occurred sensor events to the registered listeners.

To ensure that a non-logged in user is not able to receive sensor updates, the `SensorService` should either prevent the non-logged in user from registering sensor listeners or from receiving sensor updates. More specifically, at least one of the following two access controls should be applied:

- At registration and switch time: the `SensorService` should allow only current users to register listeners to receive sensor events and should unregister all listeners belonging to a user once he is no longer logged in.
- At dispatch time: the `SensorService` should deliver sensor events only to listeners belonging to the current user.

To investigate if any of these access control points is enforced, we have developed an app that logs detected sensor event changes over a specific threshold under the following two scenarios:

- The app schedules registration to receive sensor events when the victim user is logged in.
- The app registers a listener to receive sensor events when the attacker user is logged in.

Similar to the media recording attack (Fig. 3), we launch the app from the attacker account and switch to the victim account. Our app is able to receive sensor events about detected changes under the victim account environment in both the two scenarios, without his awareness. The logs are saved in the attackers account. Please note that we have performed the attack for all sensors available on a Google Nexus 7 tablet. The attack is successful in all cases.

From the results of our experiment, we can infer that none of these access control points are enforced. To confirm this, we investigate how apps register listeners to receive sensors events and how events are delivered.

An app registers a listener to receive sensor events through invoking `SensorManager.registerListener`. We checked the call chain of this API and found out that the `SensorService` does not apply any access control to check if the app registering a sensor listener belongs to the current user, which explains why our app was able to register a listener under the first scenario. Moreover, by inspecting the `SensorService` code, we found out that it does not unregister listeners belonging to non-current users once a user switch takes place.

However, no access control at registration/switch time does not necessarily imply a security flaw if proper access control is enforced at dispatch time. However, the success of our attack proves that there is no access control at dispatch time based on user status. We confirm this by inspecting `SystemSensorManager.DispatchSensorEvent` call chain, and finding that the system delivers sensor events to all listeners regardless of the status of the user who registered them.

*3) Location sensor:* We have performed the same attack on the GPS location sensor and found that a non-current user cannot succeed in getting GPS location updates of the logged in user. We investigated `LocationManagerService` and found that it applies proper access control based on user status at dispatch time. Specifically, the method `LocationManagerService.handleLocationChangedLocked` will only dispatch location updates to the registered listeners belonging to the current user, using the `userId` check shown here:

```
int receiverUserId = UserHandle.getUserId(receiver.mUid);
if (receiverUserId != mCurrentUserId) {
    // skip update
    }
```

Here, the `LocationManagerService` keeps track of the current user in `mCurrentUserId`. Every time a user switch happens, `LocationManagerService` changes its value to reflect the id of the new logged in user. A similar design should be applied to other shared hardware resources.

### D. Shared Package Information

Hypotheses 4 and 5 state that specific problems may occur due to the fact that Android apps belonging to different users share critical package information:

1) Apps sharing the same `appId` in different users share permissions. As a result, the effective permission of these apps is the union of the declared permissions for each app and the `sharedUserId` apps escalate their permissions. This is the essence of Hypothesis 4.
2) An app installed for different users share the same app package information. Consequently, one user may trigger a package update to modify the app's manifest file or code without other users' consent. This is Hypothesis 5.

To design an experiment to confirm these two problems, we need to first understand more about how the package manager stores and uses an installed app's package and its relevant information. In the package manager, all the package information among users are stored in a global hash map `mPackages`.

The keys of this hash map are package names, and the values are packages including permissions and code information of the packages. Hence, we realize that `mPackages` is app name-based, rather than user-based, confirming the platform-centric approach to package management that remains at the core of the multi-user framework. With this as a basis for our understanding, we can now discuss the testing of each of these hypotheses separately.

*1) Permission leakage in* `sharedUserId` *apps*: Android's `sharedUserId` feature allows apps signed with the same key to share permissions and data. Previous work in the single user environment has shown this convenience feature to have risks due to implicit capability leaks among apps [11]. Although `sharedUserId` app's data ends up being properly isolated in multi-user due to Linux's use of the `uid` (which accounts for both `appId` and `userId`), this is not the case with permissions. In fact, these capabilities are leaked across user boundaries, even if a particular user only has one of the `sharedUserId` apps installed. This occurs because of the platform-centric design of `PackageManager`.

During installation, permission sets are stored in `packages.xml`, while installation status is stored in separate `package-restrictions.xml` files for each user. For `sharedUserId` apps, permissions from each app are combined within the `<shared-user>` block in `packages.xml`. During boot, the package manager loads this permission list into the hash map `mPackage` in a way that makes it impossible to separate the individual permissions from each `sharedUserId` app in case a particular user does not have them all installed. As a result, when `sharedUserId` apps from the same developer are installed in varying combinations by different users on the same device, every single app gains the union of permissions from all of the `sharedUserId` apps installed on the platform, regardless of which `sharedUserId` apps have been installed by that particular user.

To confirm this, we created a pair of `sharedUserId` apps. `shareduidapp1` declares `INTERNET` permission, and `shareduidapp2` declares `READ_CONTACTS` permission. We then installed `shareduidapp1` under the owner's account only, and `shareduidapp2` under a secondary account only. After installation, we observed the following snippets present in `packages.xml`:

```
<package name="com.example.shareduidapp1" ... sharedUserId="
    10056">
<package name="com.example.shareduidapp2" ... sharedUserId="
    10056">
<shared-user name="com.example" userId="10056">
    <perms>
        <item name="android.permission.READ_CONTACTS" />
        <item name="android.permission.INTERNET" />
    </perms>
```

Here, `shareduidapp1` and `shareduidapp2` share `userId` 10056 as shown in the `<package>` blocks above. Separate from the package names, within the `<shared-user>` block, `userId` 10056 is then associated with the two permissions. However, no structure retains the fact that the `INTERNET` permission was contributed by `shareduidapp1`, and `READ_CONTACTS` was contributed by `shareduidapp2`. The record of which users have these apps installed is contained in each user's `package-restrictions.xml` file.

Because of the commingling of permissions within `packages.xml`, when user 10 runs `shareduidapp2`, the system grants both `INTERNET` and `READ_CONTACTS` permissions even though `shareduidapp1` is not installed for the user. Meanwhile, Settings reports `shareduidapp2` only holds `READ_CONTACTS` permission. This condition also occurs for `shareduidapp1` run by user 0. Each user is unaware of the permission leakage and over-privilege. Moreover, if user 10 is a restricted profile for which the owner carefully enabled apps based on their reported permissions, this leakage could allow the restricted profile accesses they should not have.

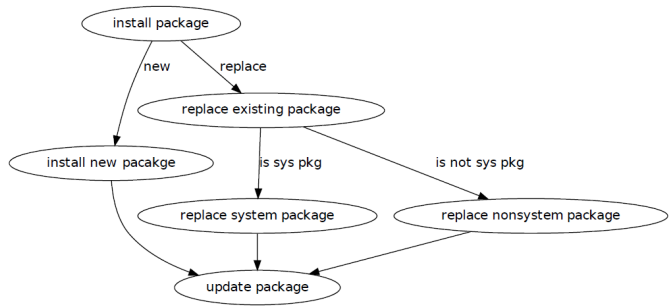

Fig. 4. Package install procedure.

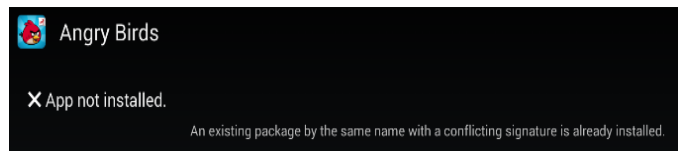

Fig. 5. New package installation is denied due to existing package with the same name but different signature.

*2) Package-based code sharing across users:* All users share the same package code for a specific package name, because the package manager creates only one package object for a specific app. To differentiate users who have installed the app, the package manager maintains an array variable called `userState`. Based on this design, there is no way for a user to have a stand-alone copy of code for a specific package.

The app install or update procedure is depicted in Fig. 4. When `PackageManager` receives the package install/update request, it will first check if this is a new install or not. As long as the package has been installed by at least one user, the package manager considers that as a update. If it is a new install, a new package object is created and the app is installed for the install user. Otherwise, the new package will replace the existing package for the install user. Before the package manager takes action to update package information, it performs a signature comparison check. If the signatures do not match, the system will deny the update request as shown in Fig. 5.

If installing an app for a specific user, the package manager will traverse all the available users and mark the installing user's `userState` variable appropriately. In special cases where an app is installed for all users, then all the users' `userState` variables are marked as installed. Since all users share the same app package information, a user can intentionally upgrade and even replace other users' packages. The most significant security impact is two-fold:

First, one user may escalate the permissions of apps belonging to a second user. For example, the latest version of Twitter requires an extra permission, `READ_SMS`, compared to the old version. The owner may choose not to upgrade to the latest one for privacy concern. However, a secondary user just updates the app through Google Play because she likes the new features. As a result, this update event will update all users' apps (if installed) without other users' consent. The newly updated package requests more permissions and performs different computing logic than the old one. In this scenario, a secondary user grants a new permission to Twitter on behalf of all the users instead of just herself.

Second, a user may have a chance to affect other users' app installation by creating denial of service (DoS) attacks in two ways. First, a user can fake a package to issue a DoS on package installation by installing a fake Facebook app before other users install the legitimate one. In such a case, no one else can install the legitimate app anymore, and no one else can uninstall the faked package through the user interface. Only the owner can force uninstalls using `adb`.

Another negative side-effect of all users sharing the same `appId` as mentioned in Section II, is that one user may use up all the `appId` values which prevents other users from installing any apps. We confirmed this by installing 50,000 dummy apps on a Nexus 10 running KitKat 4.4 as a secondary user, thus using up all available `appIds`. As a result, any other user including the owner cannot install apps any more. The Android log shows that the installation failure is because `INSTALL_FAILED_INSUFFICIENT_STORAGE`, but actually there is still space in data partition. The failure is because all users share the same `appId` range.

The root cause for code sharing problem is that Android does not provide clear code separation for different users. All the users share the same package information including app code, `appIds`, and their privileges for installing apps is mixed together. The package manager fails to isolate the code space of each user, although this design significantly saves the valuable disk space.

## V. RELATED WORK

Our work is inspired by those who have reported on confused deputies [12], [13], [14], [15], [16], [17], component hijacking [18], and capability leaks [11], [19] in Android. However, these focus on maliciousness among *apps*, while our work addresses this among *users*. Several studies [20], [21], [22], [23] have explored the use of motion sensors available on smartphones to perform user activity recognition, while [24], [25], [26], [27], [28] have focused on inferring user keyboard presses, icon taps and secure inputs using accelerometer and gyroscope sensor. Nonetheless, these do not consider unauthorized use of these sensors by other users of the device as we do. Finally, [29] and [30] present complete secure multi-user architectures which may have application in solving some of the problems we have pointed out.

## VI. CONCLUSION

We have described the basics of multi-user support in Android and outlined a systematic approach to studying whether Android's security model is properly adapted to this new environment. Our investigation methodology does not begin with a particular attack, threat model, or vulnerability, but instead seeks to provide insights into potential problems through a comprehensive analysis of all subject-object combinations. With these insights, we are able to put forth a number of hypotheses related to specific attacks. We describe the experiments designed to test these, and the findings for each. Our results indicate that several of our hypothesized concerns are in fact true. Some findings, such as those related to `MediaServer`, may be relatively easy to address by adding additional access control logic at key locations. Others however, such as ones stemming from Android's approach to package management, are architectural issues which will require system design changes. Since we believe it to be inevitable that Android will continue its expansion into environments where user trustworthiness cannot be relied upon, we see a clear need to continue the systematic investigations we describe.

ACKNOWLEDGEMENT

We would like to thank the anonymous reviewers for their valuable and encouraging comments. This work was supported in part by NSF grants 1017771 and 1318814, and AFRL project GAIHCYBR. Any opinions, findings, conclusions or recommendations expressed in this material are those of the authors and do not necessarily reflect the views of the NSF or the US Air Force.